\documentclass[letter,traditabstract]{aa}
\usepackage{graphicx}
\usepackage{txfonts}

 \newcommand{\mic}{$\mu$m}

\def\cone {\ifmmode{{\rm C}{\rm \small I}(^3\!P_1\!-^3\!P_0)}
     \else{C\ts {\scriptsize I}{\small$(^3\!P_1\!-^3\!\!\!P_0)$}}\fi}
\def\ctwo {\ifmmode{{\rm C}{\rm \small I}(^3\!P_2\!-^3\!P_1)}
     \else{C\ts {\scriptsize I}{\small$(^3\!P_2\!-^3\!\!\!P_1)$}}\fi}

\def\tex {\ifmmode{{T}_{\rm ex}}\else{$T_{\rm ex}$}\fi}
\def\tmb {\ifmmode{{T}_{\rm mb}}\else{$T_{\rm mb}$}\fi}
\def\ci     {\ifmmode{{\rm C}{\rm \small I}}\else{C\ts {\scriptsize I}}\fi}
\def\hi     {\ifmmode{{\rm H}{\rm \small I}}\else{H\ts {\scriptsize I}}\fi}
\def\hh     {\ifmmode{{\rm H}_2}\else{H$_2$}\fi}

\def\ts     {\thinspace}
\def\kms    {\ifmmode{{\rm \ts km\ts s}^{-1}}\else{\ts km\ts s$^{-1}$}\fi}
\def\msol   {\ifmmode{{\rm M}_{\odot}}\else{M$_{\odot}$}\fi}
\def\lsol   {\ifmmode{{\rm L}_{\odot}}\else{L$_{\odot}$}\fi}
\def\zsol   {\ifmmode{{\rm Z}_{\odot}}\else{Z$_{\odot}$}\fi}
\def\etal   {{\rm et\ts al.}}

\def\wateru  {{\rm H$_2$O}($J_{K_{a}K_{c}}$=1$_{10}$$\to$1$_{01}$)}
\def\waterd  {{\rm H$_2$O}($J_{K_{a}K_{c}}$=2$_{11}$$\to$2$_{02}$)}

\begin{document}

\title{Discovery of an extremely bright submillimeter galaxy  at z=3.93}
 \subtitle{}

\author{
J.-F. Lestrade\inst{1}
\and
F. Combes\inst{1}
\and
P. Salom\'e\inst{1}
\and
A. Omont\inst{2}
\and
F. Bertoldi\inst{3}
\and
P. Andr\'e\inst{4}
\and
N. Schneider\inst{4}
}
   \institute{Observatoire de Paris, LERMA, CNRS, 61 Av. de l'Observatoire,
              F-75014, Paris, France
              \and
              Institut d'Astrophysique de Paris, UMR 7095, CNRS, UPMC Univ. Paris 06,
              98bis Boulevard Arago,  F-75014, Paris, France
              \and
             Argelander Institute for Astronomy, University of Bonn, 
            Auf dem H\"ugel 71, 53121 Bonn, Germany
              \and
              Laboratoire AIM Paris-Saclay, CEA/IRFU/SAp - CNRS -,
             Universit\'e Paris Diderot, 91191 Gif-sur-Yvette Cedex, France }

   \offprints{J-F. Lestrade, \email{jean-francois.lestrade@obspm.fr}}

   \date{Received 2 September 2010; accepted 13 October 2010}
   \titlerunning{An extremely bright SMG at z=3.93}

   \abstract{Serendipitously we have discovered a rare, bright submillimeter
galaxy (SMG) with a flux density of 30 $\pm$ 2 mJy at $\lambda=1.2$mm, using MAMBO2 
at the IRAM 30-meter millimeter telescope. 
Although no optical counterpart is known for MM18423+5938, we were able to measure the redshift $z=3.92960 \pm 0.00013$ from the detection of CO lines 
using the IRAM Eight MIxer Receiver (EMIR).  
In addition, by collecting all available photometric data in the far-infrared and radio to constrain its spectral
energy distribution, we derive the  FIR luminosity 4.8 10$^{14}/m$ \lsol \ and mass  6.0~10$^9/m$~M$_\odot$  
for its dust, allowing for a magnification factor $m$ caused by
a probable gravitational lens. The corresponding  star-formation rate is 8.3 10$^{4}/m$ \msol/yr. 
The detection of three lines of the CO rotational ladder, and
a significant upper limit for a fourth CO line, allow us to estimate  an \hh\,  mass of between 1.9 10$^{11}/m$ M$_\odot$ 
and 1.1 10$^{12}/m$ M$_\odot$. The two lines \cone\, and \ctwo\, were clearly detected and  yield  
a  [\ci]/[\hh] number abundance of between 1.4 10$^{-5}$ and 8.0 10$^{-5}$. 
Upper limits are presented for emission lines of 
HCN, HCO$^+$, HNC, H$_2$O, and of other  molecules. The moderate excitation of the CO lines
is indicative of an extended starburst, and excludes the dominance of an AGN in heating this high-z SMG. 
}


\maketitle

\section{Introduction}
Deep blank-field millimeter and submillimeter  surveys of small fields ($\sim$ 1 deg$^2$)
have revealed many dusty,  starburst submillimeter galaxies (SMGs) over the past decade
with flux densities of a few to about ten mJy at  $\lambda=1.2$mm    
({\it e.g.}  Bertoldi et al. 2007, Greve et al. 2008), and higher at $\lambda=850\mu$m owing to dust
emissivity ({\it e.g.} Smail et al 1997). Recently, the South Pole Telescope  
survey, less deep but much larger in sky coverage  (87~deg$^2$), has found 47 brighter SMGs with  
flux densities  between 11 and 65 mJy at  $\lambda=1.4$mm (Vieira et al. 2010). 
Whereas redshifts of  SMGs are crucial to study  their physical properties, most of these dust-obscured galaxies 
have very faint or no optical counterparts, making measurements of  spectroscopic redshift 
extremely difficult or impossible ({\it e.g.} Smail et al. 2002). 

These  dust-enshrouded star-forming galaxies are expected to be at high  redshifts and 
are identified with the most massive galaxies assembled during  an energetic early phase of galaxy formation. 
Their abundance appears  to peak at $z \sim 2.5$  (Chapman et al. 2005, Wardlow et al 2010). 
Their star formation rate is prodigious at up to  10$^3$ M$_{\odot} yr^{-1}$, and the underlying 
starburst activity is believed  to result from mergers (Blain et al. 2002).  

Lestrade et al. (2009) discovered  serendipitously a  rare, bright point source, 
MM18423+5938, at  $\lambda=1.2$mm (30 mJy) by mapping 50 separate fields
totalling a  sky area of 0.5 deg$^2$ with the MAMBO2 bolometer camera (Kreysa et al at the IRAM 30-meter millimeter telescope.
{ Subsequently, some of us (PA and NS)  searched but did not find  local CO in the direction of the source suggesting not 
a young stellar object but an SMG instead, despite 
the Vieira et al's  cumulative source count that yields a  chance as low as $\sim$ 7\% of finding a 30mJy SMG.}  
MM18423+5938 is detected at 70~$\mu$m but undetected at  24~$\mu$m in MIPS/Spitzer images.  
It is in neither the 2MASS catalogue, nor
the NVSS VLA catalogue ($S_{1.4GHz} <$ 2.5mJy), and no optical and X-ray identifications are found in
catalogues searched with  NED (MM18423+593 is however outside the SDSS footprint). 
All these photometric data are summarized
in Table \ref{tab:continuum}. 

We show in this Letter  that  MM18423+5938 is a
bright, high-redshift SMG.  
We present in Sect. 2 our IRAM/EMIR spectroscopic observations  of MM18423+5938
at millimeter wavelengths that  yielded  our detections of CO and \ci, 
and upper limits for other molecular species. 
In Sect. 3, we model  the data to infer the dust and gas content of MM18423+5938 and its general properties.  
To compute distances and luminosities, we adopt the $\Lambda$-CDM concordance
cosmological model,  H$_0$ = 71 km/s/Mpc, 
$\Omega_{\small M} $ = 0.27, and $\Omega_{\Lambda}$ = 0.73 (Hinshaw et al. 2009).

\section{Observations and data analysis}

 Lestrade et al. (2009) detected  MM18423+5938 with a high but uncertain integrated 
flux density of $30-60$ mJy, given that it was located   
close to the border of their  MAMBO map. We reobserved 
MM18423+5938 with MAMBO at the IRAM 30-m telescope 
in the  on-off  wobbler-switching mode  on 2010 January 16  using the map  coordinates 
$(\alpha_{2000}=18^h 42^m 22.5^s \pm 0.2^s$ and $\delta_{2000}=59^{\circ} 38' 30'' \pm 2'' )$ and    
 measured  $30\pm 2$ mJy  at $\lambda=1.2$mm.

   \begin{table}[!b]
      \caption[]{Photometry available for MM18423+5938}
         \label{tab:continuum}
            \begin{tabular}{l c c }
            \hline
            \noalign{\smallskip}
            Band &  $S_\nu$ (mJy) & Ref \\
            \noalign{\smallskip}
            \hline
            \noalign{\smallskip}
 1.4 GHz   &  $<$ 2.5     &  NVSS  \\
  3 mm & 2$^{+2.0}_{-1.5}$  &  this work \\
  2 mm & 9$\pm$3  &  this work \\
  1.2 mm & 30$\pm$2  &  this work \\
 100\mic  & $<$600   &  IRAS \\
 60\mic  & $<$100   &  IRAS \\
 70\mic  &  31$\pm$4   &  Spitzer/MIPS \\
 24\mic  &  $<$ 0.6  &  Spitzer/MIPS \\
  B V R I   & B mag $ >$ 21  &  NED (DSS) \\
            \noalign{\smallskip}
            \hline
           \end{tabular}
\end{table}

\begin{figure}[!b]
\resizebox{7.75cm}{!}{\includegraphics[angle=0]{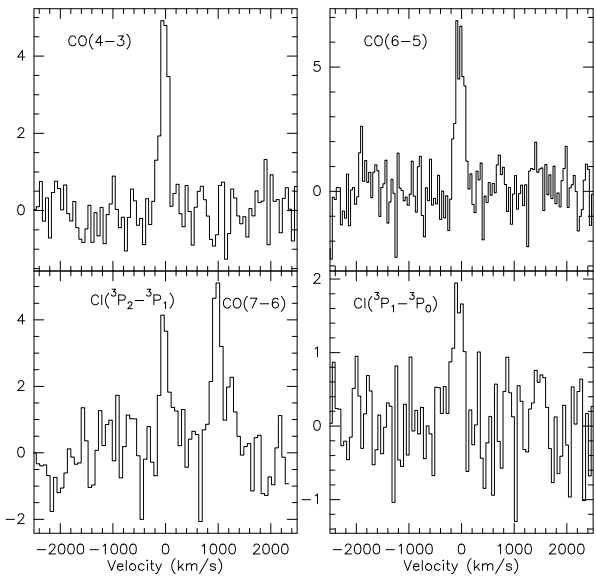}}
\caption{The three CO rotational lines detected, and the two \ci\, lines.
The vertical scale is T$_{mb}$ in mK.
}
\label{fig:spectra}
\end{figure}

MM18423+5938 is located close to the border of the archived MIPS/Spitzer maps 
centered on the star GJ725AB (AOR 4199424). We determined the 
flux densities of the star GJ725AB and  MM18423+5938 at  70~$\mu$m and 24 ~$\mu$m by means of 
aperture photometry,
scaling with the stellar photospheric flux densities  of GJ725AB  predicted by  the NextGen stellar atmospheric 
model (Allard et al. 2001).  We note a difference 
of 9'', {\it i.e.} at the $3 \sigma$ level, between 
the positions of the source in our MAMBO map (Lestrade et al. 2009)
and the  archived 70~$\mu$m MIPS Spitzer map, which is not understood.

To measure the redshift  of  MM18423+5938, we used the strategy of observation developed by Weiss et al. (2009) 
with  the multi-band heterodyne receiver Eight MIxer Receiver (EMIR \footnote{http://www.iram.es/IRAMES/mainWiki/EmirforAstronomers})
at the IRAM 30-m telescope.
The 3mm setup (E090) of EMIR provides 7.43~GHz of instantaneous, dual linear 
polarization  bandwidth. 
The entire frequency range from 77.7 to 115.8~GHz in the 3mm band can be searched with six 
tunings spaced to provide 0.5~GHz overlap.    
This range corresponds to $0 < z < 0.48$ and  $1 < z < 10$ for the CO lines between (J=1-0) and (J=8-7).
Observations were conducted from 2010 July 29th to August 2nd with precipitable waper vapor 
comprised between 3 and 7mm 
and with standard system temperatures of 110K for the E090 setup. Data were
 processed with  16 units of the Wide band Line
Multiple Autocorrelator (WILMA)  providing a spectral resolution of 2 MHz for the E090 setup.    
The observations were conducted in wobbler-switching mode, with a switching frequency of 1~Hz and an azimuthal
wobbler throw of $100''$. Pointing and focus offsets were determined once every two hours  
and  found to be stable. Calibration
was done every 6 minutes using the standard hot/cold load absorber. The data were reduced with the CLASS software.

We started to scan the whole 3mm band by integrating data for $\sim$ 2~hrs  for each tuning. 
We discovered unambiguously a 
line at 93.52 GHz after 20 minutes of integration during our third tuning on the second night, and  
continued to integrate dual polarisation 
data for 1.5~hr in total to reach an rms noise level of T$_{mb}$=0.8 mK in 60\kms\, channels
(Fig~\ref{fig:spectra} lefthand top panel).  At this stage, we successively assumed that this line  
could be CO(1-0), CO(2-1), ... to calculate for each of these assumptions the corresponding redshift 
and predict the frequencies of the  higher $J$ transitions
 accessible in the 2mm band of EMIR (setup E150 from 127 to 176~GHz). We then tuned to these 
frequencies with the E150 setup 
and swiftly detected a line at 140.26~GHz that corresponds to  CO(6-5) at  $z=3.92960 \pm 0.00013$, 
in addition to the line at  93.52~GHz for CO(4-3). 
This identification of the CO transitions and determination of the redshift of  MM18423+5938 were carried out 
 during the same night of July 30/31 in $\sim$ 6 hours.
The rest of the allocated time (15 hours) was used to search for CO(7-6) (detected), CO(9-8) 
(undetected, consequently CO(10-9) was not searched), 
and for other  species, \ci\, (two lines detected),
and HCN \& HNC(5-4), LiH(1-0) \& HCO+(5-4), H$_2$O$_o$, H$_2$O$_p$, 
$^{13}$CO(5-4), CS(8-7), and CS(9-8)  (all undetected but interesting
upper limits are discussed below, see Table \ref{tab:limits}).   Two CO lines (5-4 and 8-7) unfortunately are in the
atmospheric O$_2$ and H$_2$O lines opacity domains and could not be observed.
The spectra were of high quality,  stable, and flat. Their mean levels measure the continuum flux densities at 2mm and 3mm 
(Table \ref{tab:continuum}) owing to the excellent weather conditions (precipitable water vapor  $\sim$ 4 mm). 

Figure \ref{fig:spectra} displays the three CO lines detected, along with
the two \ci\, lines. Spectra were smoothed to a resolution of 30-50 km/s. 
Gaussian models were fitted to the lines and the  results are reported in Table
\ref{tab:lines}. Line widths found are normal {\it albeit} small suggesting that we are observing a galaxy
seen rather face-on.  

\begin{figure}[!t]
\resizebox{6.75cm}{!}{\includegraphics[angle=0]{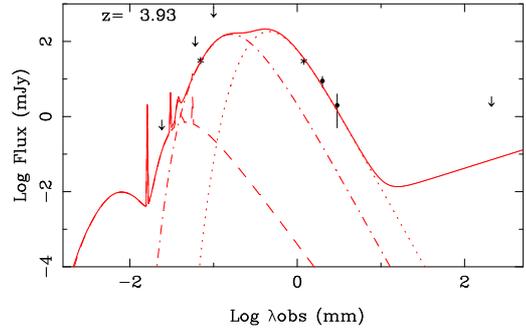}}
\caption{Available photometric data for MM18423+5938 (Table \ref{tab:continuum}). These are 
superimposed upon our dust emission model (full red curve)  based on the  Milky Way dust model by
 Desert et al (1990) adapted for our source at z=3.93. It consists of three
components ;  the large grains at T$_d$=45K, containing most of the mass (dotted line), the very
small grains at hotter temperature (dot-dash), and the PAH (dashed line).
}
\label{fig:SED}
\end{figure}

   \begin{table*}
      \caption[]{Observed line parameters}
         \label{tab:lines}
            \begin{tabular}{l c c c c c c c}
            \hline
            \noalign{\smallskip}
            Line & $\nu_{\rm obs}$ & \tmb & $S_\nu$ & $\Delta
      V_{\rm FWHM}$ & $I$& $V^{*}$ & $L'$/m/10$^{10}$             \\
                 & [GHz] & [mK] & [mJy] & [\kms] & [Jy \kms]&[\kms] & [K  \kms\,
pc$^2$] \\
            \noalign{\smallskip}
            \hline
            \noalign{\smallskip}
 CO(4--3)       & 93.5249    &  5.3  $ \pm$ 0.6 &  26.7 $ \pm$ 3.     & 175 $\pm$ 
20   &
                   4.95 $\pm$ 0.5 &  $0$ $\pm$ 8 &19.7$\pm$2.  \\
 CO(6--5)       &  140.2695   &  6.4  $ \pm$ 1&  32.8 $ \pm$ 5.     & 189 $\pm$  19 
 &
                   6.6 $\pm$ 0.6 &  $-7$ $\pm$ 9&11.5$\pm$1.  \\
 CO(7--6)       &  163.6342   &  4.2  $ \pm$ 0.9&  21.5 $ \pm$ 4.6     & 172 $\pm$ 
27   &
                   3.9 $\pm$ 0.5 &  $11$ $\pm$ 11 &5.0$\pm$0.6  \\
\cone             &  99.8378 & 1.9 $ \pm$ 0.6 &  9.6 $\pm$ 3.  & 225 $ \pm $ 55
 & 2.3 $ \pm$ 0.5  & $-50$ $\pm$ 24 &8.0$\pm$1.7 \\
 \ctwo           &   164.1802 & 4.2 $ \pm$ 1. &  21.5 $ \pm$ 5.     & 184 $ \pm $ 49
 & 4.2 $ \pm$ 0.8  & 8 $\pm$ 19 &5.3$\pm$1. \\
            \noalign{\smallskip}
            \hline
           \end{tabular}
\begin{list}{}{}
\item[] Quoted errors are statistical errors from Gaussian
            fits. The systematic calibration uncertainty is 10\%
\item[$^{*}$] The velocity is given relative to z=3.929605.
\end{list}
\end{table*}

\section{Results}

\subsection{Dust emission}

The photometric data are collected in Table \ref{tab:continuum}.  
{ To model these data in  Fig.\,\ref{fig:SED}, we choose to use the well-established Milky Way dust model 
of Desert et al (1990) as a template. This model consists of three main components, PAH and both very small and large grains, and
the emissivity slope is assumed to be $\beta$=2.
The large dust grains are dominant in mass and their temperature is estimated to be T$_d$ = 45K, constrained by the 
Rayleigh-Jeans part of the emission.   
The very small grains are made of  two temperature components,  $\sim$ 80 and  $\sim$ 130K,  constrained
 by the 70$\mu$m  Spitzer measurement and the robust relationship between $\rm L'_{CO(3-2)}$ and $\rm L_{FIR}$ 
found by  Iono et al (2009). 
A single temperature component for the small grains in the model was tried but the resulting $\rm L_{FIR}$
is significantly inconsistent with this relationship.}
The total mass of the dust required by this model is M$_{dust}$=6.0~10$^9/m$~M$_\odot$, and for a
gas-to-dust mass ratio of 150, we infer that M$_{gas}$=9.2~10$^{11}/m$~M$_\odot$. 
We suspect  there is a gravitational lens along the line of sight, with an amplification factor $m$, 
or these values would be implausibly  at least one order of magnitude larger than for  rare hyperluminous objects.
We derive the total FIR luminosity L$_{\rm FIR}$ = 4.8 10$^{14}/m$ \lsol \, 
and the star-formation rate SFR = 8.3 10$^{4}/m$ \msol/yr, by applying the relation of Kennicutt (1998).

\begin{figure}[!t]
\centering
\resizebox{4.5cm}{!}{\includegraphics[angle=-90]{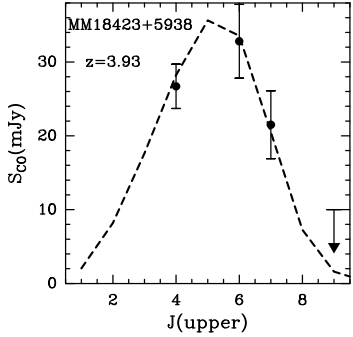}}
\caption{CO line flux density (points) or upper limits (arrow) measured
at the IRAM-30m, with the best-fit LVG model (dash line)
computed with a density n$_{\hh}$ = 10$^3$ cm$^{-3}$, T$_{k}$ = 45K,
{ and N$_{\rm CO}$/$\Delta$V = 3 10$^{18}$ cm$^{-2}$/\kms.}}
\label{fig:SLED}
\end{figure}

\subsection{CO lines}

The CO SED of MM18423+5938 in Fig. \ref{fig:SLED} indicates moderate line excitation, peaking only at J=5;
CO SEDs  peak at J=6 or 7 in local starbursts such as  M82 and NGC253 (Weiss et al 2007), peak at higher J 
in AGN-dominated sources,  {\it e.g.} J=10 in APM0827, and  reach a plateau  for $\rm J > 8$ in Mrk231 
(van der Werf et al 2010).
   We ran several LVG models, to constrain the \hh\, volumic density
and the kinetic temperature (cf Combes et al 1999). The moderate excitation implies
a regime of low temperature and/or density. 
The gas kinetic temperature is taken to be equal to the dust temperature (T$_d$=45K).
When models are run with higher T$_k$, they all imply   n(\hh) lower than 10$^3$cm$^{-3}$, which is
not realistic for CO(7-6)-emitting clouds. Our estimate is therefore
T$_k$=45K, n(\hh)=10$^3$cm$^{-3}$, { and a column density per velocity interval
N$_{\rm CO}$/$\Delta$V = 3 10$^{18}$ cm$^{-2}$/\kms,}
as adopted to model the data in Figure \ref{fig:SLED}.
From the CO(1-0) line derived from our LVG model, we  infer
an \hh\,  mass of 1.9~10$^{11}/m$ M$_\odot$ with the 
conversion factor M(\hh)/L'$_{\rm CO}$=0.8 M$_\odot$/(K \kms pc$^2$) adopted
for ULIRGs by Solomon et al (1997). This latter value yields a lower limit, while the standard
conversion ratio for the Milky Way, 5.75 times higher, yields the upper limit
M(\hh)=1.1~10$^{12}/m$~M$_\odot$. The true mass must lie between these two values, and indicates
large amounts of molecular gas even if allowing for an amplification $m$.
Assuming a typical intrinsic extension of 3kpc (e.g. Tacconi et al 2006), the surface filling factor 
of the molecular component is 0.3 for $m$=1, and 0.03 for $m$=10. 
The L$_{\rm FIR}$/L'$_{\rm CO}$ of 2~10$^3$~\lsol/K~\kms\,~pc$^2$ 
is consistent with  ratios of other luminous infrared galaxies within the scatter (Iono et al 2009).


\subsection{Atomic carbon lines}

The two lines \cone\, and \ctwo\, were clearly detected.
They have comparable central velocity
and line width (Table \ref{tab:lines}), implying that they originate 
from the same region in the source.
The relation between the
integrated \cone\, brightness temperature and the beam averaged \ci\, column density
with { the usual assumption of the optically thin limit} is given by
$$
{\rm N}_{\ci} = \frac{8 \pi k \nu_{10}^{2}}{h c^3
A_{10}}\,Q(\tex)\,\frac{1}{3}\,{\rm e}^{T_{1}/\tex} \int \tmb\,dv,
$$
 where $Q(\tex)=1+3{\rm e}^{-T_{1}/\tex}+5{\rm e}^{-T_{2}/\tex}$ is the \ci\,
 partition function, and $T_{1}$\,=\,23.6\,K and $T_{2}$\,=\,62.5\,K are the
 energies above the ground state.
When dealing with high-z sources, we can use 
 the definition of the line luminosity
 (e.g. Solomon \etal\, 1997) and derive the \ci\, mass via
(cf Weiss et al 2003, 2005)
$$
{\rm M}_{\ci} = 1.902\times10^{-4}\,Q(\tex)\,{\rm e}^{23.6/\tex}\,L'_{\cone} [\msol].
$$
 The mass estimated from the higher-excitation line is 
expressed in an analogous way, and we can then deduce that
$$
\frac{L'_{\ctwo}}{L'_{\cone}}  = 2.087\,{\rm e}^{-38.9/\tex}\,,
$$
where line luminosities are given in Table \ref{tab:lines}.
The derived  \tex \,  is 33.9 K. The mass of atomic carbon thus amounts to
M$_{\ci}$=1.0~10$^8/m$~M$_\odot$.
Given our lower and upper limits to the \hh\, mass from the CO lines, 
 the [\ci]/[\hh] number abundance is between 1.4 10$^{-5}$ and 8.0 10$^{-5}$.
This is somewhat  higher than the average   [\ci]/[\hh]  number abundance found in 
comparable star-forming objects 
(e.g. Barvainis et al 1997, Pety et al 2004, Weiss et al 2003, 2005, Riechers et al 2009, Danielson et al. 2010).
In these latter estimates, although the observed L$_{\ci}$/L$_{\rm CO}$ values are comparable, the [\ci]/[\hh] 
number abundances are somewhat dissimilar
because of the various CO-to-\hh\, conversion factors adopted by these authors.
Abundance lower than 1.8 10$^{-5}$  has
been reported (Casey et al 2010).
The contribution of the atomic carbon to the cooling is low,
L$_\ci$/L$_{\rm FIR}$ = 2.5 10$^{-6}$, comparable to that of nearby star-forming 
galaxies (Gerin \& Phillips 2000).

   \begin{table}[!t]
      \caption[]{Line upper limits}
         \label{tab:limits}
            \begin{tabular}{l c c c}
            \hline
            \noalign{\smallskip}
            Line &  $\nu_{obs}$   &  $S_{\nu}  (3 \sigma)$  & $L'$/10$^{10}$     \\
                 & [GHz]          & [mJy]                   & [K  \kms\,pc$^2$]  \\
            \noalign{\smallskip}
            \hline
            \noalign{\smallskip}
 CO(9--8)  &  210.344   &   $<10.$ &  $<$1.4\\
 $^{13}$CO(4--3) &  89.412 &  $<10.$ & $<$8.1\\ 
  HCN\& HNC(5--4) & 90.  &  $<7.$ & $<$5.5\\
  LiH(1--0) \& HCO+(5--4) &90.  & $<7.$ & $<$5.5 \\
   H$_2$O$_o$(1,1,0-1,0,1) &  112.978 &  $<10.$ & $<$5.0\\
   H$_2$O$_p$(2,1,1-2,0,2)&  152.554 &  $<12.$ & $<$3.3\\
  CS(8--7)    & 79.488  &  $<11.$ & $<$11.2\\
  CS(9--8)    &  89.420 &  $<7.$ & $<$5.6\\
            \noalign{\smallskip}
            \hline
           \end{tabular}
\end{table}


\subsection{Other lines}

We  searched for high-density tracers, such as HCN, HNC, and HCO$^+$,
in particular their lowest level available, i.e. J=5-4. The upper limits found
(Table \ref{tab:limits}) confirm that, on average, the \hh\, density is not high, as found by
our LVG models of the CO line excitation. The HCN luminosity is
higher than one third of the CO luminosity in local AGN-dominated objects 
(Imanishi et al 2004), and we note that our $3 \sigma$ upper limit  
($\rm L'_{HCN(5-4)}/L'_{CO(5-4)} \sim 0.28$) is close to this limit.
However, observation of  HCN(1-0) and CO(1-0) are needed to conclude.

We  also searched for H$_2$O emission, using 
the first ortho and para lines in their ground states, 
i.e. \wateru\, and \waterd\, but only
obtained interesting upper limits.  Assuming that
these water lines are optically thick,
these upper limits yield a filling factor of
dense clumps lower than 30\% that of CO clouds.

At z=2.3, a tentative detection of the \waterd\, line 
was reported in IRAS F10214 (Encrenaz et al. 1993;
Casoli et al. 1994), while at z=0.685, the fundamental transition of ortho-water,
\wateru\ , was  detected in absorption towards B0218+357 (Combes \&
Wiklind 1997).
Water lines were searched for other starburst galaxies
at high z (Riechers et al., 2006,
Wagg et al.2006), and significant upper limits set.
A search for \wateru\ emission toward the $z$=3.91 quasar
APM\,08279+5255 provided a surface-filling factor lower than 12\%
of the CO one (Wagg et al. 2006, Weiss et al. 2007).

Finally, other interesting molecules were undetected in the observed bands, although
with insufficient sensitivity:
 $^{13}$CO(4--3) (limiting the $^{13}$CO/$^{12}$CO
emission ratio to $<1/2$),  CS(8--7), LiH (1--0) lines,
several  lines of formaldehyde H$_2$CO, and  SiO, which is a tracer of shocks. 
Their lowest transitions observed are  $\rm H_2CO(7(2,5)-8(0,8))$,~  
$\rm H_2CO(6(1,6)-5(1,5))$,~  SiO(9-8) and upper limits are $\rm L' < 20~10^{10}$ K  \kms\,pc$^2$.   

\section{Discussion and conclusion}

We have found that the brightest SMG in the North, MM18423+5938,  is at a redshift $z=3.92960 \pm 0.00013$.
This source is part of a most interesting population of similar objects recently found by
the Herschel Atlas survey (Negrello et al 2010 submitted). 
From our modelled SED, the FIR luminosity 4.8~$10^{14}/m$~\lsol \ and mass  6.0 10$^9/m$ M$_\odot$ 
for the dust  of  MM18423+5938, 
and the implied  star-formation rate of 8.3~10$^{4}/m$~\msol/yr, lead to suspect 
a gravitational lens along the line of sight with an amplification factor $m$.
From our modelled CO~SED, we have found that M(H$_2$) is between 1.9
and 10.8~10$^{11}/m~$M$_\odot$ depending on the conversion ratio, indicating large amounts of molecular gas. 
The CO line excitation is moderate which indicates that there is both no strong heating by a central AGN and
 a starburst that is not too extreme. The average density of the molecular
medium is low, of the order of 10$^3$ cm$^{-3}$, and the gas kinetic temperature is assumed to be 45K in our model. 
The atomic carbon lines, assumed optically thin, 
are excited to \tex~=~33.9K and yield  M$_{\ci}$=1.0~10$^8/m$~M$_\odot$, which corresponds to
a  [\ci]/[\hh] number abundance of between 1.4 10$^{-5}$ and 8.0 10$^{-5}$ when the limits on the  \hh\, mass
 derived from the CO lines are used. 
In this high-z SMG,  the \ci-to-CO luminosity ratio is consistent with those of other high-z galaxies.

The moderate CO line excitation found excludes a dominant AGN in MM18423+5938,
unlike Mrk231 where CO is excited up to J=13 (van der Werf et al 2010).
This moderate excitation favors
an extended gas disk (typically 3kpc), rather than a compact nuclear starburst (300pc)
and consequently a high CO-to-\hh\, conversion ratio.
This high-redshift SMG, with a star formation efficiency of 
L$_{\rm FIR}$/L'$_{\rm CO}$=2400, is comparable to the lower-z submillimeter
galaxies studied by Greve et al (2005).

\begin{acknowledgements}
Based on observations carried out with the IRAM 30m telescope. IRAM is
supported by INSU/CNRS (France), MPG (Germany) and IGN (Spain). 
The authors are grateful to  the IRAM staff for their support, and to the referee for helpful comments.
\end{acknowledgements}







\end{document}